\documentclass[runningheads]{svmult}

\usepackage{makeidx}   
\usepackage{graphicx}  
\usepackage{subeqnar}  
\usepackage{multicol}  
\usepackage{physprbb}  
\makeindex             

\begin{document}
\title*{
  CROMOS: A cryogenic near-infrared, multi-object spectrometer for the VLT}
\toctitle{
  CROMOS: A cryogenic near-infrared, multi-object spectrometer for the VLT}
\titlerunning{
  CROMOS: A cryogenic near-infrared, multi-object spectrometer for the VLT}
\author{R. Genzel\inst{1}
  \and R. Hofmann\inst{1}
  \and D. Tomono\inst{1}
  \and N. Thatte\inst{1}
  \and F. Eisenhauer\inst{1}
  \and M. Lehnert\inst{1}
  \and M. Tecza\inst{1}
  \and R. Bender\inst{2}
}
\authorrunning{R. Genzel et al.}

\institute{Max-Planck Institut f\"ur extraterrestrische Physik,
  85741 Garching, Germany
\and
  Universit\"ats-Sternwarte der Ludwig-Maximilians-Universit\"at M\"unchen,
  81679 M\"unchen, Germany
}

\maketitle

\begin{abstract}
We discuss a cryogenic, multi-object near-infrared spectrometer as a
second generation instrument for the VLT. The spectrometer combines 20
to 40 independent integral field units (IFUs), which can be positioned
by a cryogenic robot over the entire unvignetted field of the VLT
($\sim$7').  Each IFU consists of a contiguous cluster of 20 to 30
pixels (0.15 to 0.25'' per pixel). The individual IFUs have cold
fore-optics and couple into the spectrograph with integrated
fibers-microlenses. The spectrometer has resolving power of
$\lambda$/$\Delta$$\lambda$$\sim$4000 and simultaneously covers the J-, H-,
and K-bands with three HAWAII 2 detectors. The system is designed for
operation both in seeing limited and MCAO modes. Its speed is
approximately 3500 times greater than that of ISAAC and 60 times greater
than NIRMOS (in H-band). The proposed instrument aims at a wide range of
science, ranging from studies of galaxies/clusters in the high-z
Universe (dynamics and star formation in z$>$1 galaxies, evolution of
ellipticals, properties of distant, obscured far-IR and X-ray sources),
to investigations of nearby starbursts, star clusters and properties of
young low mass stars and brown dwarfs.
\end{abstract}

\section{Motivation and Science Drivers}
Optical photometry/spectroscopy is the easiest and most commonly used
technique for studying stellar and interstellar components in galaxies.
Yet extinction/reddening by dust and cosmological redshift are two of
the reasons that make observations in the near-infrared (NIR: $\lambda$
$\sim$ 1--2.4 $\mu$m) necessary and/or highly attractive. The much
better performance of adaptive optics at infrared wavelengths is
another. Furthermore the NIR emission traces older stellar
populations that are better measures of stellar mass in galaxies, and
there are a number of NIR spectral features (e.g. H$_2$, CH$_4$,
H$_2$O etc.) that uniquely trace cool interstellar and circumstellar
gas. Because of poor detector performance and size, low instrument
transmission and high sky brightness, however, infrared observers until
a few years ago had to pay a price of $>$3 mag in sensitivity compared
to optical spectroscopy, making NIR spectroscopy of distant
or faint sources challenging or impossible (Figure \ref{fig2}). With
the advent of new high quality detectors and spectrometers, such as ISAAC on
the VLT and NIRSPEC on the Keck telescope
\cite{McLean1998}\cite{Moorwood1998}, combined with software suppression of the
OH sky emission lines, NIR long-slit spectroscopy has become
competitive with optical spectroscopy. Adaptive optics assisted,
integral field spectroscopy (e.g. with SINFONI on the VLT
\cite{Thatte1998}) will soon open up sensitive, near-diffraction limited,
NIR imaging spectroscopy. The next obvious steps are
NIR multi-object spectroscopy with cryogenic slit masks.
Two examples are
FLAMINGOS for GEMINI \cite{Elston1998} and LUCIFER for the LBT
\cite{Mandel2000}. An additional option is a cryogenic multi-object
spectroscopy with independent integral field units. Although it is
certainly a challenging endeavor, it is
also the most versatile and sensitive option. It is this option that
we propose here. 

As an example of the wide range of science issues that can be addressed
with such an instrument, some of the key science drivers are,

\begin{enumerate}
\item\label{proj1} dynamics and physical characteristics of z$>$1 star
  forming galaxies,
\item\label{proj2} evolution of the most massive galaxies, notably
  ellipticals at z$>$1
\item\label{proj3} redshifts and properties of distant dusty starbursts
  and faint hard X-ray AGNs,
\item\label{proj4} age dating/population studies of starbursts and
  stellar clusters, and
\item\label{proj5} spectroscopy of young low mass stars and brown dwarfs.
\end{enumerate}

Project \ref{proj1}) calls for high quality H$\alpha$ (and other
emission lines)
profiles in a large number of galaxies of different characteristics, for
studying the cosmological evolution of galaxy mass and
studying the cosmological evolution of galaxy mass and mass-to-luminosity
ratio.  Spatially resolved measurements with $\ll$ 0.5''
resolution (in good seeing, or with AO) are necessary to resolve
rotation curves in sources that are typically 1'' to 2'' in diameter.
Both hierarchical merger scenarios in a cold dark matter
Universe and current observations at z$<$1 and z$>$2.5 indicate that 
for 1$<$z$<$2
galaxy properties should exhibit rapid
evolution, which would be most pronounced for the most massive systems.
J/H/K observations are necessary for addressing the H$\alpha$ line. Another
goal is the investigation of metallicities, stellar populations and star
formation histories through emission line ratios, where spatially
resolved information is desirable but not in all cases necessary (or
possible). In conjunction with recent theoretical models, studies of the
dynamics and star formation properties of large samples of high-z
galaxies, especially in the critical range between z=1 and z=2.5
(reachable only or primarily with infrared observations), will give a
better understanding of the processes involved in galaxy
formation/evolution and the formation of the Hubble sequence. This
project (as well as others on the list) exemplifies the shift in
observational cosmology during the next years away from pure redshift
studies, and toward detailed spectroscopic investigations of the
physical properties of 
large samples of 
individual objects.

Project \ref{proj2}) aims at
the important issue of how the most massive galaxies evolve,
when and how elliptical galaxies were formed,
and whether some ellipticals formed very early (z$\geq$3) through major
starbursts. It will require high signal-to-noise ratio spectra to get at
key absorption lines 
(Balmer lines, Mg b, G-band, etc.) 
for redshift
determinations, measurements of velocity dispersions, 
metallicities and ages.
Distant ellipticals and dusty star bursts
are often extremely red (R$-$K$>$5) so
that only infrared spectroscopy has a chance of getting redshifts and/or
information on their physical characteristics.

Project \ref{proj3}) requires
determination of emission line redshifts for larger samples of the faint
far-infrared/submm sources found by ISO, SCUBA and MAMBO (and in the
future by SIRTF, FIRST/HERSCHEL and ALMA) and of the hard X-ray sources
currently discovered by XMM-NEWTON and CHANDRA. The final objective of
this project is an investigation of the relationship and relative
cosmological evolution of powerful starbursts and accreting massive
black holes.

The goal of project \ref{proj4}) is an empirical
determination of the temporal and spatial evolution of nearby starbursts
for a more detailed understanding of global star formation processes
(feedback, superwinds, initial mass function, globular cluster formation,
etc.). Apart from being a key issue in its own right, this project also
serves as input for the high-z studies mentioned above.

Project
\ref{proj5}) aims at a better understanding of the properties and
evolution of young (pre-main sequence) low mass stars and brown dwarfs
through detailed NIR spectroscopy. These cool objects have
characteristic molecular absorption bands in the 1--2.4 $\mu$m that can
be used, in conjunction with theoretical models, to more quantitatively
understand their structure, atmospheres and evolutionary state. 

Common to all these projects is that they aim at a physical
understanding of faint objects through detailed, spatially resolved
spectroscopy at the best possible sensitivity and with a broad
wavelength coverage of samples with statistically meaningful
sizes.

\begin{figure}
\begin{center}
\includegraphics[width=\textwidth]{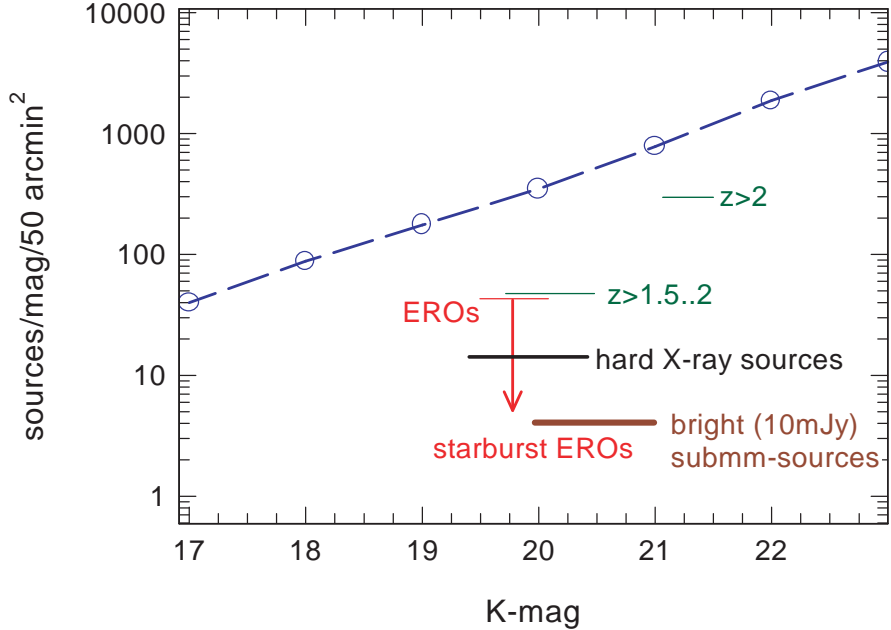}
\end{center}
\caption{
  Source surface densities as a function of K-band magnitude. The
  dashed curve and circles denotes the K-band source counts
  \protect\cite{Bershady1998}. The fractions of those sources at high
  redshift are denoted by upper bars
  \protect\cite{Ferguson2000}\protect\cite{Rudnick2001}.
  The source density of 2--10 keV sources ($\geq$2$\times$10$^{-15}$
  erg/s/cm$^2$ \protect\cite{Giacconi2001})
  is indicated as a thick bar. The surface
  density of extremely red objects (EROs: R$-$K$>$5)
  is marked by a bar \protect\cite{Daddi2000}, with an arrow denoting the
  fraction of EROs that are likely dusty starbursts (rather than
  early-type galaxies). The source density of bright (S$_{850\mu
  m}$$\geq$10 mJy)
  submm sources \protect\cite{Blain1999} is indicated by the lowest
  bar.
}
\label{fig1}
\end{figure}

A critical question is the expected density of sources. Figure
\ref{fig1} summarizes the surface densities of the sources in projects
1--3. For the magnitude limits of a ground-based instrument on the VLT
(AB$\sim$21--23, K$\sim$19--21), there are typically between a few to a
few tens of sources in the 7' unvignetted field of view of the VLT.

\section{Instrument characteristics}

To achieve the requirements just mentioned, a suitable multi-object
spectrometer has to have the following characteristics:

\begin{enumerate}
\item cryogenic operation and OH sky line suppression/avoidance for
  optimization of sensitivity out to 2.4$\mu$m,
\item simultaneous coverage of J-, H-, and K-bands at spectral resolving
  powers sufficient for high quality dynamical studies and for software
  OH suppression ($\lambda$/$\Delta$$\lambda$$\geq$4000),
\item individual integral field units covering typical object sizes
  ($\geq$2'') at pixel scales suitable for spatially resolved studies
  ($\sim$0.15'' to 0.25''), and
\item 20--40 IFUs movable over at least the unvignetted field of view of
  the VLT at Nasmyth focus.
\end{enumerate}

It is also highly desirable that such an instrument can interface with and
work behind wide-field adaptive optics systems (such as multi-conjugate
AO: MCAO), in order to image at $<$0.3'' resolution.

\begin{figure}
\begin{center}
\includegraphics[width=\textwidth]{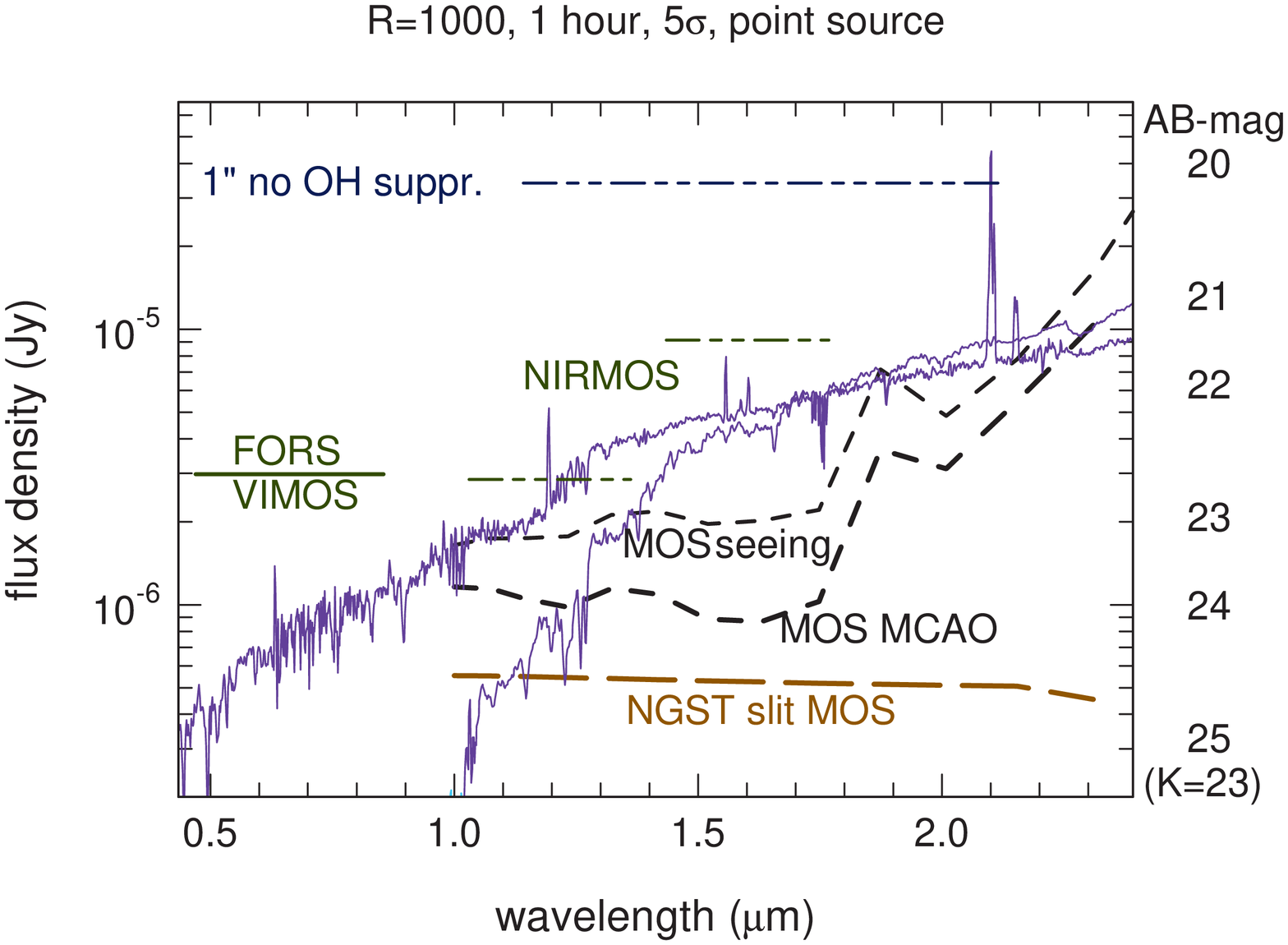}
\end{center}
\caption{
  Point source sensitivity (5$\sigma$, 1hour integration time at
  $\lambda$/$\Delta$$\lambda$=1000, upper dashes) of the proposed
  cryogenic MOS, in comparison to other instruments/facilities. For
  comparison the spectra of a moderately extincted starburst galaxy
  (bluer spectrum) and an elliptical galaxy (redder spectrum) are shown, for a redshift
  of 2.2 and an AB magnitude of 22 \protect\cite{Kinney1996}. Note the very
  substantial improvement between OH suppressing/avoiding NIR
  spectrometers with sub-arcsecond slits (such as ISAAC, NIRSPEC and
  NIRMOS) compared to instruments without these features (even on an 8m
  telescope: long-short-short dashed line). NIRMOS (long-short
  dashed bars) is not cryogenic, however, resulting in non optimum
  H-band performance (and no K-band capability). Because of its integral
  field nature, the CROMOS gains another factor of $\sim$2 for compact
  sources, since the effective `slit' can be optimized for the shape of
  the source and the seeing. The lower dashed curve
  gives the performance of the MOS in conjunction with a high
  performance (2$\mu$m Strehl ratio 0.7) MCAO system and 0.2''
  (software) pixels. Only the NGST multi-slit MOS (lower long dashes)
  outperforms the CROMOS, especially at $\lambda$$\geq$2$\mu$m.
  Between OH sky lines, the point source performance of the
  CROMOS (in terms of flux density of AB magnitude:
  AB=$-$2.5logS$_\nu$(Jy) + 8.9) is significantly better than that of
  optical multi object spectrometers, such as FORS or VIMOS. Note,
  however, that a few percent of the wavelength range covered by the MOS
  are strongly affected by OH sky lines, and thus must be discarded for
  faint source spectroscopy.
}\label{fig2}
\end{figure}

The point source sensitivity of our proposed CROMOS fulfilling the
above requirements is shown in Figure \ref{fig2}, in comparison to other
facilities/instruments and to z$\sim$2 elliptical/starburst galaxies.

\section{Key technologies}
\subsection{Spectrometer layout}
The basic layout of the spectrometer and camera derives its
design and heritage from that of the SPIFFI spectrometer for the VLT
\cite{Eisenhauer2000}. It consists of
pre-optics to re-image the object plane from the AO,
an image-slicer to cut the two-dimensional field into a set of slitlets,
and a spectrometer with an f/30 collimator and an f/1.4 camera.
Another important function of the pre-optics is to suppress the thermal
background with the cold stop.
The camera is a refractive optics using commercial glasses.
The key difference with the SPIFFI design,
other than the image slicer,
will be the separation of the
J-, H- and K-band light in the collimated part of the beam through
beam splitters, followed by three independent gratings,
cameras and detectors covering these three bands. With three HAWAII 2
detectors and $\lambda$/$\Delta$$\lambda$$\sim$4000 the entire
wavelength range from 1$\mu$m to 2.4$\mu$m can be sampled at the Nyquist rate.
The alternative of directly dispersing light onto the detectors with a single
grating is very difficult due to the challenging camera optics (f/1),
the very wide wavelength range, and due to the fact that the HAWAII 2
detectors are non-buttable. 

\begin{figure}
\begin{center}
\includegraphics[width=0.4\textwidth]{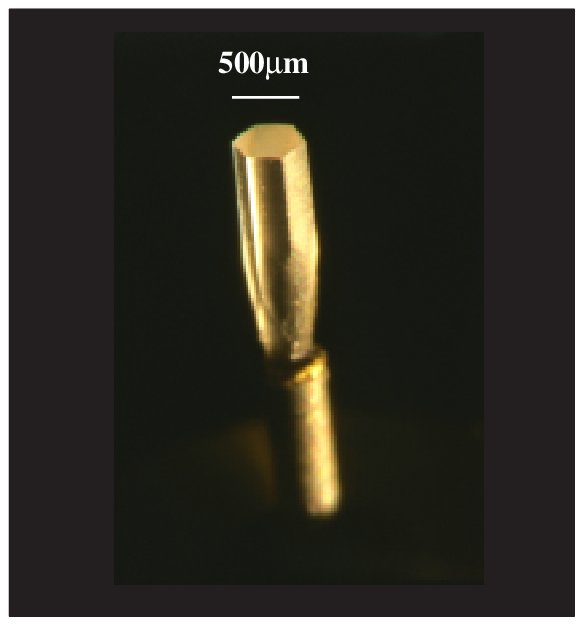}
\includegraphics[width=0.4\textwidth]{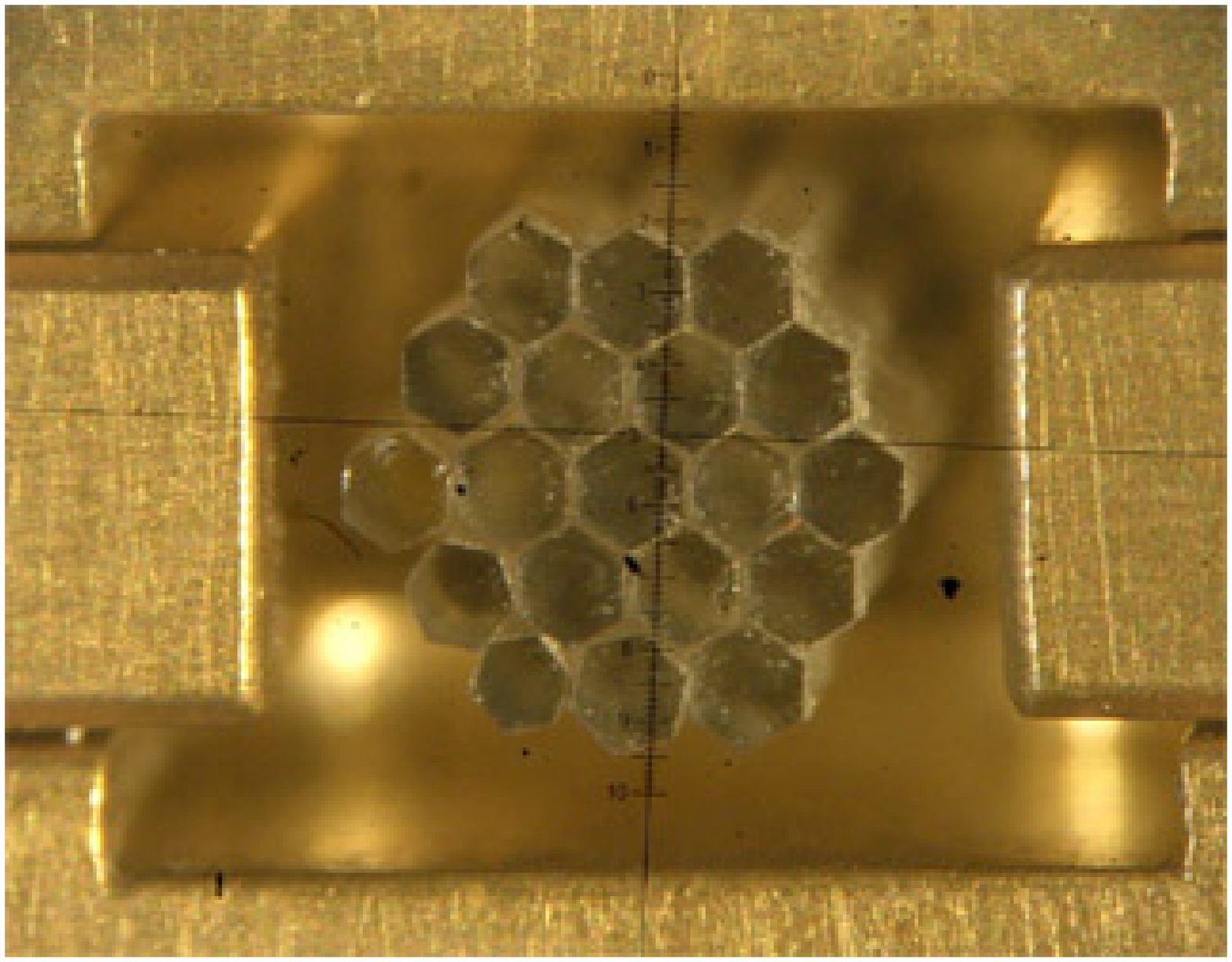}
\end{center}
\caption{
  Flared fiber-integrated microlens technology \protect\cite{Tecza1999}. The
  left inset shows a single flared fiber (core diameter
  50$\mu$m) with its integrated hexagonal microlens.
  The right inset shows a 19 element fiber IFU that is somewhat similar
  to the IFUs intended for the CROMOS.
}\label{fig3}
\end{figure}

\subsection{Integrated flared fiber-microlens IFUs}
We have tackled the well known problem of low coupling efficiency
through fibers from the telescope into a spectrograph, especially in a
cryogenic environment, by developing the `integrated flared
fiber-microlens' technology \cite{Tecza1999}. This
concept uses the polished, flared tip of a silica fiber as an integrated
microlens, thus efficiently coupling light from the telescope pupil into
the core of the fiber (left inset of Figure \ref{fig3}). The hexagonal
surface cut of the tip allows a contiguous areal coverage as required
for the IFUs of the MOS discussed here (right side of Figure \ref{fig3}).
While Figure \ref{fig3} shows a hexagonal arrangement of a fiber cluster,
other configurations (e.g. reactangular) are possible and may better
match the elongated shapes of distant galaxies. With $\sim$15 cm fiber
lengths we have achieved throughput of $>$80\% at 2 $\mu$m from an f/62 input
beam to the end of the fiber. The flared fiber-integrated microlens
technology
minimizes reflection
losses and misalignments in the cryogenic environment.
Nevertheless we are presently also
investigating the performance of such separate microlens-fiber systems.
In the CROMOS the lengths of the fibers would have to be $\sim$50 cm, thus
reducing the transmission to about 85\% at the upper part of the
K-band, near an absorption feature of silica, but little affecting the
performance at $\leq$2.4 $\mu$m. The primary challenge of our flared fiber
development up to now was the small tolerances of the
micro-machining/polishing, thus leading to poor production yields of
high throughput fibers.  This issue will have to be dealt with in the
next phase of the development.

\subsection{Spider arms and cryo-robot}
\begin{figure}
\begin{center}
\includegraphics[width=\textwidth]{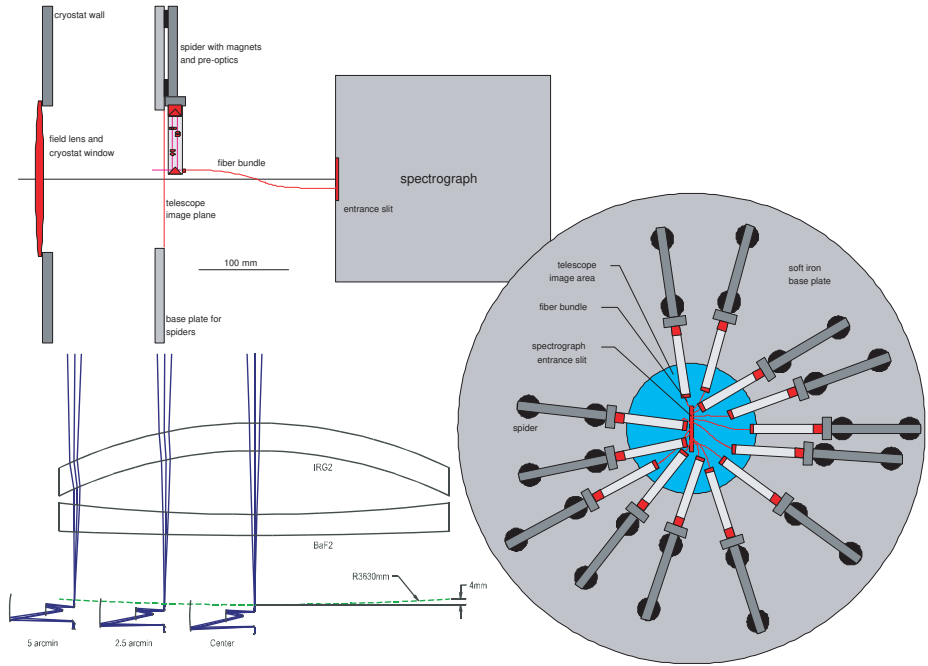}
\end{center}
\caption{
  Optical layout of the CROMOS. Top left: Side view showing field lens,
  cold fore-optics for one of the spiders mounted on the magnetic plate,
  coupling its fiber bundle into the spectrograph. Bottom left: Ray
  tracing through the field lens and fore-optics. Bottom right: Top view
  showing the `fishermen-around-the-pond' arrangement of the spiders.
}\label{fig4}
\end{figure}

Cold fore-optics are required to efficiently couple the light into a
given IFU/fiber cluster and reduce the thermal background.
For this purpose and for the mechanical
movement of the IFUs across the field of view, we are adopting a
`fishermen-around-the-pond' arrangement of spider arms. Figure
\ref{fig4} shows
the mechanical arrangement and optical trains of the spider arms
containing the fiber clusters, along with their cold fore-optics.
A field lens at the dewar window is required for centering the pupil
onto the cold stop wherever the IFU is on the focal plane. Figure
\ref{fig4} shows an example of optical design
in the bottom left. The field lens changes the direction of the light into the
right angle to the focal plane, which is spherical. The spider arms are
fastened with
magnets at the back of the spherical, soft iron plate (Figure \ref{fig4}
bottom right).

\begin{figure}
\begin{center}
\includegraphics[width=\textwidth]{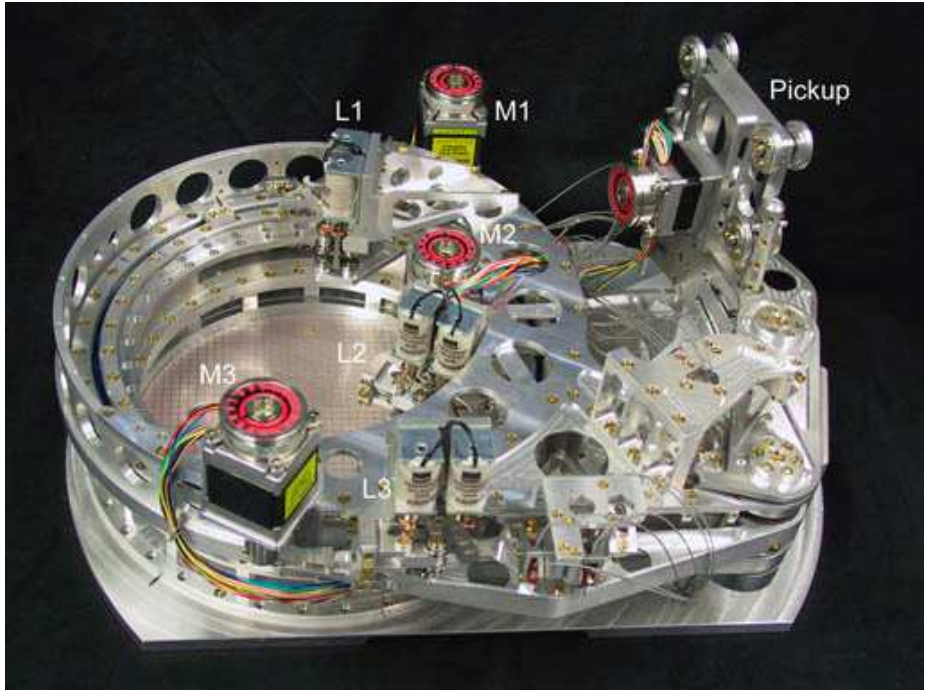}
\end{center}
\caption{
  Photograph of the cryogenic robot that places the fiber arms on the
  magnetic plate. The robot has three degrees of freedom with three
  cryogenic stepper motors (M1--3). The Pickup is coupled with a spider arm to
  be moved. Entire robot is locked by locking mechanisms (L1--3) when not
  in use. A rotation of the entire device and
  two linear movements along two sides of a parallelogram in principle
  allow placing the spider arms anywhere across the field of view,
  including rotation of the spider around the optical axis.
}\label{fig5}
\end{figure}

We have developed a cryogenic robot to place the spider arms anywhere in
the field of view. The robot (Figure \ref{fig5}) has three degrees of
freedom, one full rotation of the entire device and two linear motions
along two side of a parallelogram. The robot is designed to work at
LN$_2$ temperature and should be able to place 20 arms in about 5 to 10
minutes. We
have developed the first version of the drive software and have
successfully tested the operation of the device at room temperature.
Cold tests will follow in the next few months.

\subsection{Overall Performance}

\begin{figure}
\begin{center}
\includegraphics[width=\textwidth]{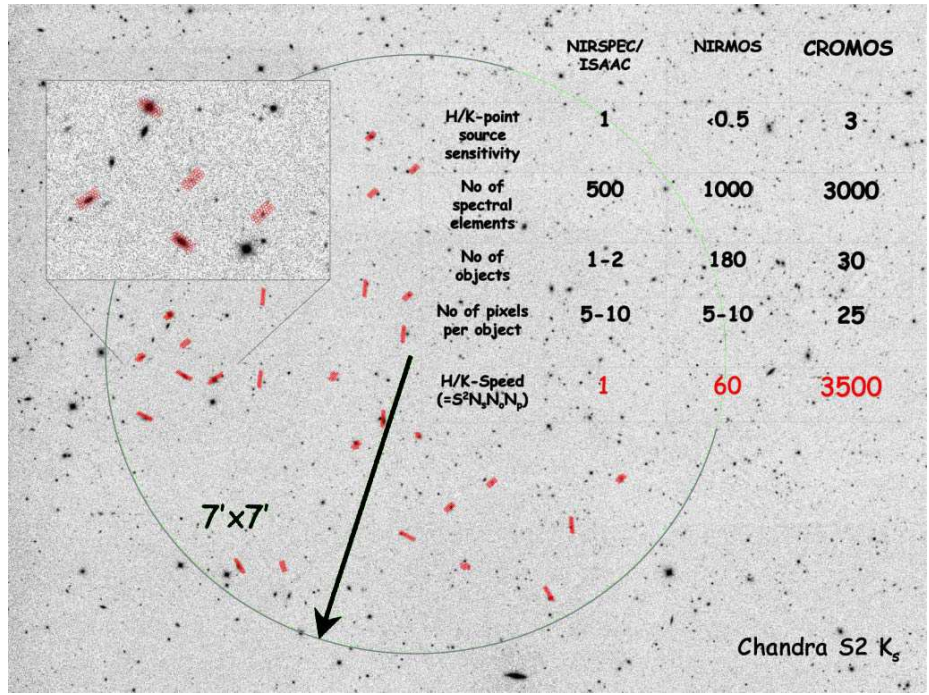}
\end{center}
\caption{
  ISAAC K$_{\mbox{s}}$ image of the Chandra South (2) area, taken as
  part of the ESO EIS Deep survey, along with a mock lay-out of $\sim$30
  IFUs from the proposed CROMOS. The table in the upper right is a
  comparison of the performance characteristics of the CROMOS, as compared
  to long slit spectrometers (ISAAC, NIRSPEC) and the warm, multi-slit
  NIRMOS instrument.
}\label{fig6}
\end{figure}

The CROMOS described above excels in H-band and K-band sensitivity,
speed (or integration time, or multiplex advantage, or sensitivity
squared),
and simultaneous wavelength coverage for detailed studies of faint
sources with a spatial extent of 1--2''. Because of its integral field
nature, the device is also optimally suited for studying faint point
sources. There are no slit losses and the source is always `on' the
detector. Figure \ref{fig6} gives an impression of a typical application
(faint galaxy studies in the Chandra South area), along with a summary
of its performance when compared to a long-slit spectrometer (ISAAC,
NIRSPEC) and a warm multi-slit MOS (NIRMOS). In terms of overall speed, the
proposed CROMOS is 3500 faster than ISAAC and NIRSPEC and about 60 times
faster than NIRMOS.

The instrumental development described here is part of a combined effort
of the Munich University Observatory and the Max-Planck Institut f\"ur
extraterrestrische Physik. It derives its overall heritage from the
SINFONI/SPIFFI and FORS developments for the VLT, and the LUCIFER
development for the LBT.

\end{document}